# Electron Photoemission in Plasmonic Nanoparticle Arrays: Analysis of Collective Resonances and Embedding Effects

Sergei V. Zhukovsky • Viktoriia E. Babicheva • Alexander V. Uskov •
Igor E. Protsenko • Andrei V. Lavrinenko

**Abstract** We theoretically study the characteristics of photoelectron emission in plasmonic nanoparticle arrays. Nanoparticles are partially embedded in a semiconductor, forming Schottky barriers at metal/semiconductor interfaces through which photoelectrons can tunnel from the nanoparticle into the semiconductor; photodetection in the infrared range, where photon energies are below the semiconductor band gap (insufficient for band-to-band absorption in semiconductor), is therefore possible. The nanoparticles are arranged in a sparse rectangular lattice so that the wavelength of the lattice-induced Rayleigh anomalies can overlap the wavelength of the localized surface plasmon resonance of the individual particles, bringing about collective effects from the nanoparticle array. Using full-wave numerical simulations, we analyze the effects of lattice constant, embedding depth, and refractive index step between the semiconductor layer and an adjacent transparent conductive oxide layer. We show that the presence of refractive index mismatch between media surrounding the nanoparticles disrupts the formation of a narrow absorption peak associated with the Rayleigh anomaly, so the role of collective lattice effects in the formation of plasmonic resonance is diminished. We also show that 5 to 20-times increase of photoemission can be achieved on embedding of nanoparticles without taking into account dynamics of ballistic electrons. The results obtained can be used to increase efficiency of plasmon-based photodetectors and photovoltaic devices. The results may provide clues to designing an experiment where the contributions of surface and volume photoelectric effects to the overall photocurrent would be defined.

S. V. Zhukovsky (✉) • V. E. Babicheva • A. V. Uskov •
A. V. Lavrinenko
DTU Fotonik – Department of Photonics Engineering, Technical University of Denmark,
Ørsteds Pl. 343, DK-2800 Kgs. Lyngby, Denmark
e-mail: sezh@fotonik.dtu.dk

A. V. Uskov • I. E. Protsenko
P. N. Lebedev Physical Institute, Russian Academy of Sciences,
Leninskiy Pr. 53, 119333 Moscow, Russia
Advanced Energy Technologies Ltd,
Skolkovo, Novaya Ul. 100, 143025 Moscow Region, Russia

V. E. Babicheva
Birck Nanotechnology Center, Purdue University,
1205 West State Street, West Lafayette, IN, 47907-2057 USA

## 1 Introduction

Incorporating of plasmonic nanostructures into the design of photodetectors and photovoltaic cells can influence the operation of these devices in many ways. For example, diffraction- or scattering-assisted light trapping and redistribution can increase the effective propagation length of a photon in a photosensitive material (e.g., semiconductor), increasing coupling of incident light to device and, in turn, the photodetection efficiency. The enhancing effects on the photodetection efficiency can be achieved in resonant structures or nanoantennas, where the excitation of localized surface plasmon polaritons can also boost optical absorption in photosensitive material of the device [1-4] and/or enhance the photoelectric effect from metal into semiconductor [5-8].

Even more intriguingly, plasmonic nanoparticles can bring about new physical mechanisms of generating photoelectrons in semiconductor-based devices. For example, in the case when the incident photon energy is below the semiconductor band gap and is therefore insufficient to generate an electron-hole pair in the semiconductor directly, introducing metal-semiconductor interfaces would result in Schottky barriers that electrons can still overcome [5-15]. Then, absorption of an incident low-energy photon by a nanostructure exhibiting a localized plasmonic resonance (LPR) with subsequent emission of a photoelectron from the particle opens an additional photocurrent generation channel [5-10,12-17]. As a result, the sensitivity range for photodetectors can be extended below the semiconductor band gap into the infrared range, limited only by the Schottky barrier height (the work function at a metal/semiconductor interface).

This effect (the "hot" electron generation) was recently observed experimentally [5-7,9-11], and is a subject of intense interest [17]. However, the detailed theoretical understanding of photoemission from nanoparticles remains largely phenomenological. Even though the general underlying physical principles were uncovered decades ago [18-25], many questions relating these general principles to the specific nanostructure geometry are still open. For example, it remains to be determined whether the energy transfer from the photons to the electrons occurs predominantly at the semiconductor/metal interface or in the bulk of the nanoparticles [19,24,25]. The role of the Schottky barrier geometry also remains to be clarified.

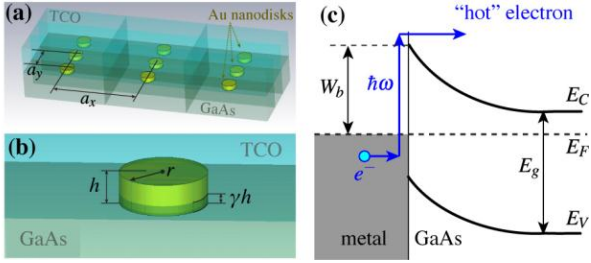

**Fig. 1** (a) Schematics of the nanodisk array partially embedded into a GaAs substrate and covered with a TCO layer. (b) Enlarged view of a single nanodisk showing the embedding geometry. (c) Schematics of the Schottky barrier at the Au/GaAs interface.

A straightforward way to elucidate these questions would be to compare the photoelectric response of nanostructures with varying geometry. Such comparison was recently done by Knight et al. [10] using metal strip grating partially embedded in a semiconductor host, with varying degree of embedding. It was shown experimentally that embedding increased the photocurrent far more strongly than the increase of the contact surface between semiconductor and metal. This anomalous increase, attributed to the momentum conservation relation between photons and electrons, suggests that photoemission should be very sensitive to nanostructure shape.

On the other hand, the arrangement of individual nanostructures into a lattice brings about collective effects and can also significantly influence the plasmonic resonance, as well as the local electric field distribution [26-31]. In particular, strong dipole interactions in a periodic array were shown to result in local optical fields orders of magnitude stronger than for isolated nanoparticles of the same size and shape [32]. It was experimentally demonstrated [33] that the quality factor of the transmission resonance in a plasmonic nanoantenna array becomes larger by a factor of 30 compared to the quality factor of a localized plasmon resonance of a single particle. The effect can be utilized to enhance performance of devices such as sensors [34] and light sources [35,36]. Recently, photocurrent enhancement in silicon diode because of an additional Fano-like resonance caused by diffractive coupling in the periodic array of nanoparticles placed on top of the device was experimentally demonstrated [37].

One recent example is the work of Sobhani et al. [11], again based on metal-strip gratings, where narrow-band photodetection can be spectrally tuned by varying the grating pitch. Another example is our earlier work [38] involving an array of nanoparticles completely embedded in semiconductor. It was shown that interaction between a narrow-band lattice resonance, namely the Rayleigh anomaly, and a broader-band localized surface plasmon resonance of individual particles brings about a tunable, narrow-band, Fano-shaped peak in spectral photoemission response.

In this paper, we build up on these earlier results and investigate the effects of partial embedding of a nanoparticle array in a semiconductor material. In this geometry (see Fig. 1), nanoparticles are connected both to semiconductor (where the Schottky barrier is formed and photoelectrons are emitted) and to transparent conductive oxide (TCO, through which the electron supply in the particles can be replenished, and through which nanoparticles can be illuminated). Combined with relative ease of fabrication, this makes partially embedded particle array a very promising geometry. However, it is shown that a refractive index step at the TCO/semiconductor interface does put a limit on maximum photoemission enhancement, lowering the optimal value of the lattice period. However, highly tunable narrowband photodetection in the nanodisk array is shown to be achievable in a broad range of geometrical configurations.

It is also shown that the degree of embedding plays an important role in tailoring the absorption and photoemission spectra of the nanoparticle array. Even though the overall photocurrent does not undergo much change in most cases, the photoemission enhancement ratio (the ratio between photocurrent generated by an embedded particle and an unembedded one) can reach the values of several tens.

The remainder of this paper is organized as follows. Section 2 describes the geometrical details of structures under study, and Section 3 reviews the theoretical background on plasmon-assisted photoelectron emission from nanoparticles. Section 4 follows with the numerical simulation results presenting the absorption and photoemission spectra of nanoparticle arrays with varying lattice period and embedding depth. Finally, Section 5 summarized the paper.

**2 Partially embedded plasmonic nanoparticle arrays**

We consider a rectangular array of metallic (gold Au) nanodisks with radius $r$ and thickness $h$, placed on top of a semiconductor (gallium arsenide GaAs) substrate with refractive index $n_m$ and covered on the other side with a TCO layer, such as indium tin oxide (ITO), with refractive index $n_t$ (Fig. 1). Further, let $a_x$ and $a_y$ denote the lattice constants in the $x$- and $y$-direction, respectively, and let $0 \leq \gamma \leq 1$ denote the embedding fraction in such a way that out of the total disk thickness $h$, the fraction $\gamma h$ is inside the semiconductor and the remaining fraction $(1-\gamma)h$ is surrounded by the TCO.

We assume that light of the frequency $\omega$ is incident normally from the TCO side, with $\omega$ satisfying
$$W_b < \hbar\omega < E_g, \qquad (1)$$
where $E_g$ is the bandgap for the semiconductor matrix, and $W_b$ is the work function for the metal/semiconductor interface. If $\hbar\omega < E_g$, no photocurrent is generated in semiconductor due to photon absorption in the band-to-band transitions. Nevertheless, if $\hbar\omega > W_b$, photocurrent can result from photoemission of "hot" electrons from metal nanoparticles into the semiconductor matrix across the Schottky barrier (Fig. 1). For gold nanodisks in a GaAs matrix, the band gap energy is $E_g = 1.43$ eV and the work function is $W_b \sim 0.8$ eV (with image force correction). Thus, Eq. (1) is fulfilled for $\hbar\omega$ from 0.8 to 1.43 eV, corresponding to a wavelength range between 870 and 1550 nm.

Each single nanodisk exhibits a localized surface plasmon resonance (LSPR) at a certain frequency $\omega_{LSPR}$ determined mainly by material indexes and the disk shape and size. If the frequency of the incident light $\omega$ is close to $\omega_{LSPR}$, the local fields inside and near the nanodisks are resonantly increased, the disks functioning effectively as plasmonic nanoantennas. These local fields in turn give rise to resonant plasmon-enhanced photoemission [5-8,10,16].

When nanostructures are periodically arranged, two regimes of lattice response can be distinguished by the electric field distribution pattern [31]: (i) where the electric fields have the maximal strength in between the particles, or (ii) where the electric field enhancement occurs inside and around the nanoparticles. The regime (i) can combine enhanced light trapping with low light absorption in metal nanoparticles and is promising for improving the photovoltaic efficiency of thin-film solar cells with direct band-to-band optical absorption [39]. In contrast, the regime (ii) gives rise to strong plasmon-related absorption and light scattering [31,32], and can lead to further enhancement of electron photoemission from metal nanoantennas compared to the case when such a collective lattice response is absent.

When assembled into a periodic lattice (Fig. 1a), the nanodisks form a 2D diffraction grating. In such a grating, Rayleigh anomalies (RAs) are known to occur at a series of wavelengths where higher orders of diffraction emerge (i.e., where evanescent waves corresponding to a certain diffraction order become propagating). For nanoparticles completely surrounded by a homogeneous medium with refractive index $n_s$ and under normal incidence of light, these wavelengths $\lambda_{RA}^{(m_x,m_y)}$ are determined by [40]

$$\frac{2\pi n_s}{\lambda_{RA}^{(m_x,m_y)}} = \sqrt{\left(m_x \frac{2\pi}{a_x}\right)^2 + \left(m_y \frac{2\pi}{a_y}\right)^2}. \qquad (2)$$

In dense lattices [7,16], where $a_x, a_y < 2\pi c/(n_s \omega_{LSPR})$, all of the RA wavelengths (2) are far away from the LSPR wavelength range $\lambda \sim \lambda_{LSPR}$, so resonant effects related to the individual-particle LSPR dominate (although it should be mentioned that $\lambda_{LSPR}$ in such an array is slightly different from $\lambda_{LSPR}$ of an isolated nanoparticle [41,42], this deviation is minor and will not be considered). However, if the lattice constants become larger so that $\lambda_{RA}^{(m_x,m_y)}$ approach $\lambda_{LSPR}=2\pi c/\omega_{LSPR}$, it is known that collective lattice resonances become pronounced and bring about enhanced narrow-band absorption and photoemission [38]. In particular, it makes sense to study a rectangular nanoparticle lattice with $a_x > a_y$ (Fig. 1a) where $a_y$ remains small in order to keep the lattice dense in the $y$-direction, while $a_x$ is varied up to large enough value for which the greatest RA wavelength,

$$\lambda_{RA} \equiv \lambda_{RA}^{(1,0)} = a_x n_s, \qquad (3)$$

becomes close to $\lambda_{LSPR}$. For $a_x$ significantly greater than $a_y$, a rectangular lattice can be regarded as an array of one-dimensional nanodisk chains. Such a chain array is somewhat akin to a metal strip grating, and is known to exhibit narrow plasmonic resonances. From the polarization properties of the short-range and long-range terms in the dipole radiation of the nanoparticles [26,28], stronger and narrower plasmonic resonances are expected for the $y$-polarization of incident light [38].

However, if the nanoparticle array is surrounded by different media on its sides, such as in Fig. 1a, the dipole moments of individual particles interact differently in the two half-spaces above and below the array, resulting in significant smearing of lattice resonances [29,43]. The exact character of this effect should strongly depend on the geometrical set-up, particularly the difference between the refractive indices $n_m$ and $n_t$, as well as the embedding fraction $\gamma$. Therefore, in what follows, we will focus on varying these two parameters and investigating the resulting changes in absorption and photoemission spectra.

## 3 Plasmon-assisted emission of photoelectrons

If strong local fields are present in the vicinity of the nanodisk surface, enhanced electron photoemission from the metal into the surrounding semiconductor is expected. Generally, two mechanisms of electron photoemission can be defined [8,19,25,26].

One mechanism is the absorption of a photon from the localized plasmon mode by an electron *inside* the nanoparticle, with subsequent transport to the surface and possible emission over the Schottky barrier (*the volume photoelectric effect*) [7,10,11,22,44]. For this effect, spatial field distribution of square of *complete* electric field over the entire nanoparticle is important, and the primary metric is the total energy absorbed by the material throughout the volume of the nanoparticle.

The other mechanism consists in the absorption of a light photon by an electron *as* it collides with the nanoparticle boundary with emission of the electron from metal, or, in other words, a photoelectron is emitted as a result of the direct action of the electric field at the surface (*the surface photoelectric effect*) [8,16,19-21,25,26]. In this mechanism, the most important factor is the distribution of squared *normal* component of electric field over the nanoplasmonic structure surface. There have been several accounts to develop the theory of the surface photoelectric effect. In Ref. [8], such a theory is based on classical works on photoelectric effect from a flat metal surface [19,20,24,25]; in Ref. [45], Govorov and co-authors develop their theory starting from the quantum microscopic description of non-equilibrium carrier population in a localized plasmon wave.

The question about which of the two mechanisms is more prevalent in each particular metallic structure is still open to discussion. However, the crucial role of strong field enhancement in and near the metal is undoubted, and this means that collective effects can play a very important part in plasmon-assisted photoelectron emission. For example, experimental results of Knight et al. [10] show a strong photocurrent increase when the degree of embedding of the nanowires is greater – far stronger than can be explained by pure geometrical increase of metal/semiconductor contact area. Using Eq. (3), one can estimate that for the considered nanowire grating pitch (500 nm) and refractive index of silicon substrate (3.5), the RA wavelength is around 1750 nm, close to the studied wavelength range (1300-1500 nm). Hence, gradual increase of nanowire embedding can enhance the collective effects in silicon, making the resonant enhancement of electric field more pronounced. This enhancement can contribute to the experimentally observed increase of photocurrent along with mechanism described in Ref. [10].

As was argued in earlier works [8,16], we will assume that surface effects prevail over bulk effects in small nanoparticles. Assuming that the nanoparticle size is still significantly larger than the de Broglie electron wavelength, the photocurrent from one nanoparticle caused by the surface photoelectric effect can be calculated. It is proportional to the squared normal component of the electric field, $|E_n|^2$, integrated over the surface of contact between the metal and the semiconductor [8,16,46]:

$$I_{NP}(\lambda) = C_{em}(\lambda) \oint_{disk} |E_n|^2 dS \qquad (4)$$

The proportionality coefficient $C_{em}(\lambda)$ depends on the properties of the Schottky barrier between metal and semiconductor, in particular the work function $W_b$ [8,44]. The photocurrent density per unit area of a photodetector device then equals

$$J_{device} = \frac{I_{NP}}{a_x a_y} = C_{em}(\lambda) |E_o|^2 \xi, \qquad (5)$$

where the dimensionless quantity

$$\xi = |E_0|^{-2} a_x^{-1} a_y^{-1} \oint_{disk} |E_n|^2 dS \qquad (6)$$

is the field enhancement factor relative to the incident field $E_0$. The intensity of the incident light in GaAs is $S = c n_m |E_0|^2 / 8\pi$. So, the quantum efficiency of photoemission from the nanoparticle array can be defined as $\eta = (J_{device}/e)/(S/\hbar\omega)$ and expressed as

$$\eta = \eta_0 \cdot \xi, \quad \eta_0 = \frac{8\pi\hbar\omega C_{em}}{n_m e c}. \qquad (7)$$

The coefficient $\eta_0$ can be estimated as $10^{-5} - 10^{-2}$ for the material parameters under study in the involved spectral range [8,16].

In our numerical simulations, we perform two types of calculations. First, we calculate the absorption spectra of the nanoparticle array. Since the volume photoelectric effect depends primarily on absorption inside the structures, the conditions of absorption increase are expected to correspond to the increase of volume photoelectric effect. Second, we perform direct calculations of photocurrent attributed to the surface photoelectric effects, based on Eqs. (4)-(7).

## 4 Results and discussion

In our numerical example we consider the structure similar to the one studied earlier [16,39], with the dimensions $r = 25$ nm, $h = 18$ nm, $a_y = 100$ nm, and $a_x$ varied between 100 and 400 nm. The permittivity of gold is described by the Drude model with plasma frequency $2.18 \times 10^{15}$ s$^{-1}$ and collision frequency $6.47 \times 10^{12}$ s$^{-1}$ [47]. The refractive index for the semiconductor is $n_m = n_{GaAs} = 3.6$, and the index of the TCO layer $n_t$ is varied between $n_{GaAs}$ and $n_{ITO} = 1.73$. Full-wave numerical calculations were carried out using CST Microwave Studio [48].

### 4.1 Nanoparticle array in a homogeneous medium

For reference, we begin by briefly reviewing our earlier results where the nanoparticle array is fully surrounded by GaAs [38]. The calculated absorption spectra are shown in Fig. 2.

For light polarized along the *x*-axis (Fig. 2a), the response of the array is modified only slightly when $\lambda_{RA}$ becomes close to $\lambda_{LSPR}$. Namely, the LSPR absorption peak experiences a moderate enhancement and a very slight shift of the wavelength, which reverses its direction: a blue shift changes to a red shift and then to a blue shift again as $a_x$ increases from 100 to 450 nm; the red shift occurs when $\lambda_{RA}$ transcends $\lambda_{LSPR}$. The peak width remains largely the same, so for the light polarized in the direction of the varying lattice constant $a_x$, the nanoparticle array is not very sensitive to its value. The response of the array is still dominated by the single-particle LSPR [16,39].

However, for light polarized along the *y*-axis (Fig. 2b), the lattice resonances exert a much more pronounced influence on the spectral response of the array. Absorption is seen to turn to zero at $\hbar\omega = 2\pi\hbar c/\lambda_{RA}$ (Fig. 2b, dotted lines), and a strong enhancement of the surface plasmon response is observed, marked by a sharp, narrow peak in the absorption spectra. This peak becomes asymmetric and Fano-shaped, which apparently results from the interaction of a narrow-band lattice resonance with a broader LSPR of the individual nanoparticle. The absorption peak wavelength is highly tunable, following $\lambda_{RA}$ towards lower frequencies as $a_x$ increases.

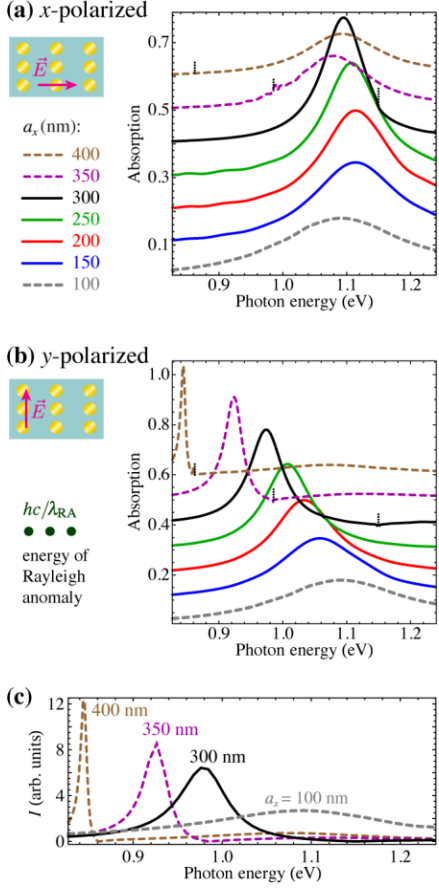

**Fig. 2** Calculated absorption spectra of nanodisk lattices with $a_y = 100$ nm and varying $a_x$ for light polarized along (a) $x$-axis and (b) $y$-axis. The dotted vertical lines on the curves for $a_x$ from 300 to 400 nm show the energy corresponding to $\lambda_{RA}$ for these $a_x$. The plots are separated by 0.1 in the $y$-axis to improve readability. (c) Spectral dependence of photocurrent from one particle $I_{NP}$ for $y$-polarized light for different $a_x$.

For the rest of the paper, we will focus on the $y$-polarized light. Using Eqs. (4)–(7) with the electric field profiles inferred form the numerical simulation, it can also be seen that sharp narrow-band absorption peak is accompanied by a corresponding peak in photocurrent $I_{NP}$ (Fig. 2c), as well as in the device photocurrent density $J_{device}$ [38]. This means that collective lattice effects in the nanoparticle array lead to narrow-band photodetection in the same way as the previously reported for metal-strip gratings [11] but for a structure containing much less metal and therefore having a much lower overall reflectivity, which may be beneficial for inclusion of the considered nanoparticle array in photovoltaic devices.

### 4.2 Effects of the refractive index step

Discrete nanoparticles have a drawback compared to an electrically connected grating: it is much more difficult to provide an electrical contact needed to replenish the emitted electrons in every nanoparticle. A common way of doing this is to cover the nanoparticle array with a TCO layer (Fig. 1a) so that part of the nanodisk surface provides the Schottky barrier and the other part provides the contact with the TCO through which the flow of electrons emitted into the semiconductor is compensated from the circuit [16].

However, as mentioned above, the refractive index step between the TCO and the semiconductor is expected to disrupt the lattice summation leading to the appearance of sharply defined RAs. Intuitively, having media with different refractive indices ($n_m \neq n_t$) to the sides of the array results in a mismatch in RA wavelengths given by Eqs. (2) and (3) for the different media, causing the lattice resonance phenomena to smear out. Even though it was shown previously [38] that these detrimental effects can be mitigated by maintaining a symmetric sandwich configuration (GaAs-TCO-GaAs), it poses a certain interest to investigate how strictly the refractive indices of the TCO and the semiconductor must match, and what are the conditions when the collective effects diminish. Furthermore, it is of interest to explore the role played by the location of the refractive index step in relation to the nanoparticles, i.e., by the embedding fraction $\gamma$.

To demonstrate these effects, Fig. 3 shows the absorption spectra for a nanodisk array on top of a GaAs substrate covered by a TCO layer with refractive index $n_t$ varying from 3.6 (which nearly equals $n_m$) down to 2.2.

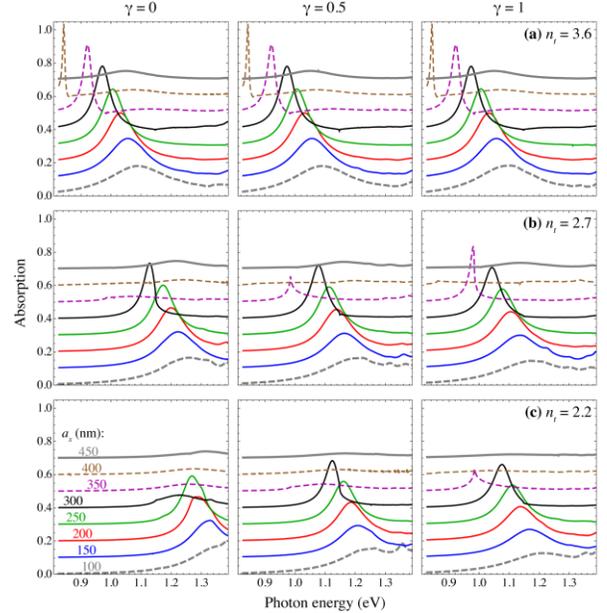

**Fig. 3** Absorption spectra for the configuration in Figs. 1a-b for (a) $n_t = n_m = 3.6$, (b) $n_t = 2.7$, (c) $n_t = 2.2$, for three values of $\gamma$ (left column: $\gamma = 0$; middle column: $\gamma = 0.5$; right column: $\gamma = 1$). The plots are separated by 0.1 for easier readability.

Several prominent features can be seen. First and most prominent, the refractive index mismatch around the nanodisk array puts a limit on how narrow the absorption peak becomes, essentially causing it to disappear once $a_x$ passes a certain threshold; the more the index mismatch, the sooner this disappearance occurs. Secondly, decreasing $n_t$ moves the LSPR towards higher frequencies, which is expected because of the overall increase of plasmonic resonance frequencies in optically less dense materials. Thirdly, the resonance structure of the LSPR becomes more complicated. This can result from a superposition of two spectrally different LSPRs of a single nanodisk in a homogeneous medium with refractive index $n_m$ and $n_t$, as well as from two distinct lattice resonances [see Eq. (3)].

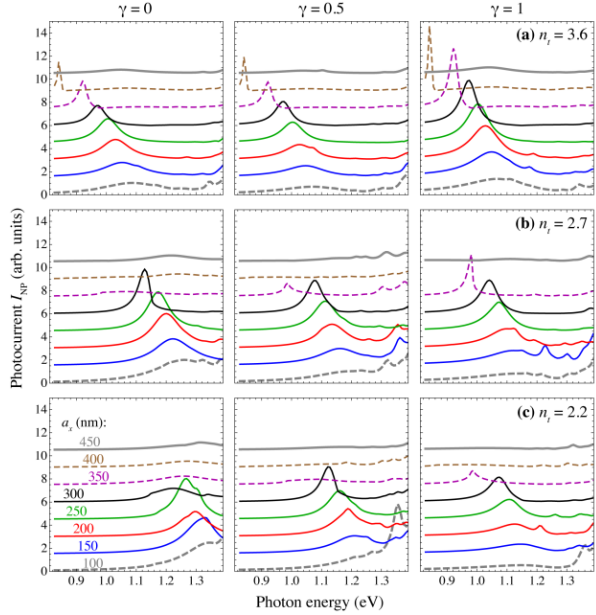

**Fig. 4** Photoemission spectra corresponding to Fig. 3. The plots are separated by 1.5 for easier readability.

It can also be noticed that the degree of embedding influences the absorption resonance rather weakly. A more embedded nanoparticle array is slightly less susceptible to the detrimental effect of the refractive index step, requiring a stronger index mismatch to cause the absorption peak to disappear. However, looking at the photoemission spectra reveals a totally different picture (Fig. 4). As regards its existence and frequency, we see that the photoemission peak follows the absorption peak, but the dependence of the peak intensity is more complicated.

When particles are more embedded in GaAs, the contact surface area between gold and GaAs significantly increases, and one would expect a corresponding increase in the photoemission current. Indeed we see such behavior for $n_t = n_m = 3.6$ (Fig. 4a). However, for different values of $n_t$, we no longer see a very significant increase in the photoemission intensity as $\gamma$ increases (except where the peak would disappear for $\gamma = 0$). This suggests that the refractive index step in the vicinity of the nanoparticle array not only destroys the collective resonance, but also diminishes the role of the sidewalls of the nanodisks in contributing to the overall photocurrent.

### 4.3 Effects of the nanoparticle embedding

As mentioned in the previous section, there is considerable freedom in choosing the embedding fraction $\gamma$ that defines the extent to which the nanoparticles are buried in the semiconductor. It was also shown (see Fig. 3) that the disappearance of collective effects in the photoemission resonance is influenced by $\gamma$, and the strength of this influence is greatly varied.

Here we analyze the effect on $\gamma$ specifically by varying it from $\gamma = 0$ (particles on top of a GaAs substrate surrounded by TCO) to $\gamma = 1$ (particles completely buried into a GaAs substrate covered by TCO), which corresponds to changing the embedding depth $\gamma h$ by 3 nm in each step. The results are shown in Fig. 5.

It can be seen that, confirming the results in Figs. 3-4, increasing $\gamma$ causes both the absorption and the photoemission resonance to shift to lower frequencies. It is also seen that the absorption peaks tend to be at their narrowest near $\gamma = 0.5$, i.e., a geometrically symmetric placement of the refractive index step makes the lattice effects strongest in this respect.

As regards the corresponding photoemission spectra (Fig. 5, bottom), we notice that they follow the same frequency shift pattern but the photoemission peak intensity can vary significantly, sometimes revealing non-monotonic dependencies (e.g. anomalously pronounced photoemission for $\gamma = 0.33$ for $a_x = 200$ nm).

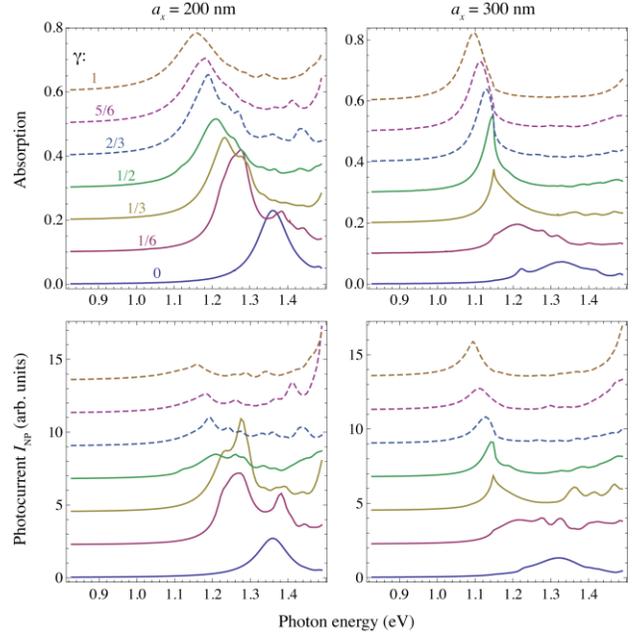

**Fig. 5** Absorption spectra (top) and photoemission spectra (bottom) showing dependence on $\gamma$ (from 0 to 1 in steps of 1/6) for $n_t = 1.73$ and two different $a_x$ (200 and 300 nm). Top plots are separated by 0.1 and bottom plots are separated by 2.25 for easier readability.

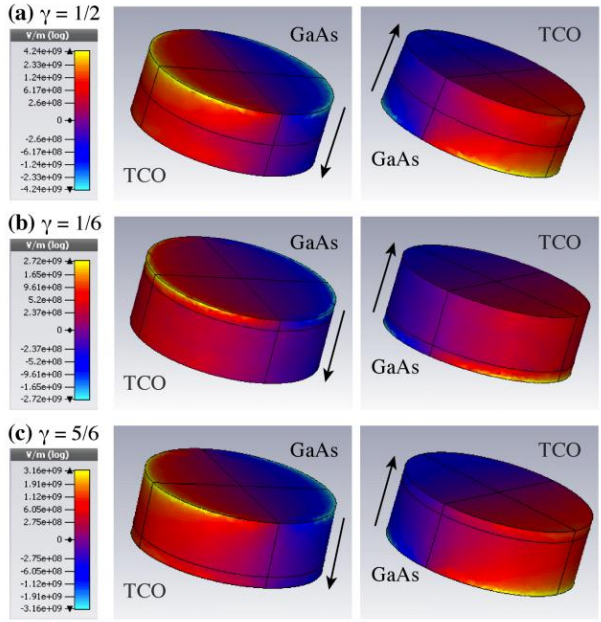

**Fig. 6** Plots of the normal field component at the nanodisk surface at the absorption peak for $a_x = 300$ nm, $n_t = 1.73$, and (a) $\gamma = 0.5$, (b) $\gamma = 1/6$, (c) $\gamma = 5/6$. Two views are shown for each $\gamma$ (from the GaAs side and from the TCO side). The arrow shows the direction of the $z$-axis (see Fig. 1).

Contrary to what was observed in Fig. 4a for $n_t = n_m$, we do not see any drastic increase of the photoemission peak as embedding increases. The reason appears to be the concentration of fields in the more optically dense semiconductor in such a way that even though the contact area of the nanodisk side wall increases, the integral in Eqs. (4-6) remains largely at the same value. Indeed, Fig. 6 reveals that the field is primarily concentrated in the semiconductor irrespective of how embedded the nanoparticles are.

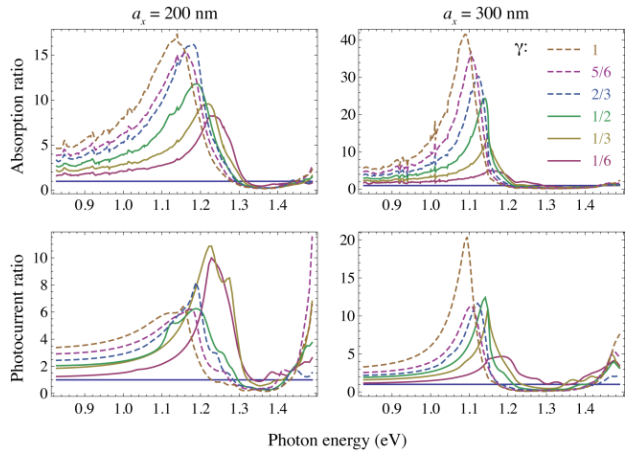

**Fig. 7** (a) Enhancement of absorption, i.e., ratios $A(\gamma)/A(\gamma=0)$ (top plots) and $I_{NP}(\gamma)/I_{NP}(\gamma=0)$ (bottom plots) over the spectral range for different periods ($a_x = 200$ and $300$ nm) and $n_t = 1.73$.

Finally, to compare the present results with those of Ref. [10], we have calculated the absorption and photocurrent ratios, $A(\gamma)/A(\gamma = 0)$ and $I_{NP}(\gamma)/I_{NP}(\gamma = 0)$, on the embedding fraction $\gamma$, over the whole studied spectrum. At some frequencies values up to 20 are reached, and values around 5 are found in the broad spectrum (Fig. 7). This effect cannot be explained by pure geometrical increase of metal/semiconductor interface. Moreover, as we did not take the electron dynamics into account, this effect cannot be ascribed to the influence of electron wavevector and matching conditions for ballistic electron emission, either. Thus we suggest that it can be an alternative or additional mechanism of experimentally observed 25-times enhancement of photocurrent respone in Ref. [10].

However, while large enhancement are seen at some frequencies, the number is large because of the small absolute value of absorption (or photocurrent) at $\gamma = 0$. As the embedding changes, the peak spectrally moves, causing large values of absorption and photoemission ratio. In terms of peak value of photoemission, we have not observed a great increase when nanoparticles become more embedded in the semiconductor, at least when the refractive index mismatch between TCO and GaAs is significant.

## 4 Conclusions

To summarize, we have studied the characteristics of plasmonic absorption and plasmon-assisted emission of "hot" electrons due to the surface photoelectric effect in metallic nanoparticle arrays partially embedded into a semiconductor (GaAs) matrix and covered by a transparent conductive oxide (TCO) layer. It is confirmed that the presence of refractive index mismatch between GaAs and TCO disrupts the formation of a narrow absorption peak associated with the Rayleigh anomaly, so the role of collective lattice effects in the formation of plasmonic resonance is diminished. In turn, this imposes a limit on the narrow-band photoemission from nanoparticle arrays reported previously [38]. This effect can be used when individual-particle effects need to be isolated from collective lattice resonances. It is also established that the degree of embedding of the nanoparticle array in the semiconductor influences this process of collective effect disruption.

The degree of embedding also has a more complicated influence on the photoemission current. When there is no refractive index mismatch netween the TCO and the semiconductor, the photocurrent is increased in accordance with the geometrical increase of the Schottky barrier area. When, however, the refractive index step is present, this geometrical effect is overridden by the electric field concentration in the semiconductor, which becomes independent of the degree of embedding (Fig. 6). As a result, the peak photocurrent becomes independent of the embedding degree, even though the total absorption increases.

However, in line with experimental results of [10] we found large values of of photoemission enhancement ratio, up to 20 times when embedding is varied by 18 nm. In contrast to the explanation put forth in [10] (the ballistic electron momentum considerations), this effect can be purely attributed to collective lattice resonances.

By supplementing the present calculation with a more involved study of the role of volume photoelectric effect, which should take into account ballistic transport considerations depending both on the initial momentum of the "hot" electron and on the geometric configuration of the Schottky barrier, one could attempt to answer the elusive question whether it is the surface effect or the volume effect which dominates photoemission in presence of collective lattice resonances. This calculation is the task for a forthcoming study.

Still, we have shown that by varying simple geometrical parameters such as the lattice constant and embedding depth, spectral properties of photoelectron emission from plasmonic nanoparticles can be varied to a great extent. This can be used for tailoring the spectral response in plasmonics-assisted infrared-range photodetectors based on nanoparticle arrays, and can find application in new types of narrow-band photodetectors and photovoltaic elements in the infrared range.

**Acknowledgments** S.V.Z. acknowledges financial support from the People Programme (Marie Curie Actions) of the European Union's 7th Framework Programme FP7-PEOPLE-2011-IIF under REA grant agreement No. 302009 (Project HyPHONE). V.E.B. acknowledges financial support from Otto Mønsteds Fond and Thriges Fond. I.E.P. and A.V.U. acknowledge support from the Russian MSE State Contract N14.527.11.0002 and from the CASE project (Denmark).